\documentclass[11pt]{book} 

\usepackage{amsmath,amsthm,amscd,amssymb}

\theoremstyle{definition}
\newtheorem*{abstract}{Abstract}

\begin{document}
\chapter{Mach's Principle}

Herbert Lichtenegger\\
Institut f\"ur Weltraumforschung\\
\"Osterreichische Akademie der Wissenschaften\\
A-8042 Graz, Austria\\

\noindent Bahram Mashhoon\\
Department of Physics and Astronomy\\
University of Missouri-Columbia\\
Columbia, Missouri 65211, USA

\begin{abstract} We briefly review the history of Mach's principle and discuss its significance in the light of modern
physics.
\end{abstract}

\section{Introduction} Whether space is an independent entity of its own, a number of relations between material objects
or a mere subjective notion superimposed on the world is a question of long tradition and probably rooted in our
common impression of the distinct difference between the objects and the empty space between them. Two different views
thus emerged from the broad range of opinions in western thinking about the nature of space and time, considering them
either as absolute or relative. The absolute view identifies space with a container holding all material objects in
which the bodies can move but which exists independently of its content, while the relative view considers space
merely as a conceptual abstraction of the storage of the bodies and is thus based on the existence of bodies,
losing its meaning without them.

Although the question about the very nature of space and time seems to be at first glance a purely philosophical
problem, it is of eminent physical significance concerning the foundation of physics since it is closely related
with the question of what constitutes the method of physics and what should be the elements of a physical theory.
Newton's concepts of absolute space and time, though having proved to be highly successful, were not accepted by
all scientists and the call for a reformulation of mechanics in terms of purely relational quantities never fell
silent. Although Mach was not the first who insisted on such a reformulation, he was without doubt the most
influential one and his critique of the Newtonian concepts of absolute space and time published in his
``Mechanics'' was later loosely termed ``Mach's principle'' by Einstein\footnote{Indeed the terms
``Machian principle" and ``Machian postulate" can already be found prior to Einstein in
Schlick \cite{schlick:1915}.} \cite{einstein:1918}. However, since Mach made only tentative proposals, there
are a great number of interpretations and different formulations of this principle \cite{barbour&pfister:1995}.

In the following section we present a brief historical review concerning the problem of absolute/relative motion.
Since Mach is well known and mostly cited among physicists within this context, we also give credit to a number of
other authors who contributed with equally interesting thoughts to the subject. For an excellent comprehensive
overview regarding various topics of Mach's principle we refer to the book edited by Barbour and Pfister
\cite{barbour&pfister:1995}.

\section{Absolute Versus Relative} The origin of the two divergent concepts of space mentioned above can be traced
back to ancient times: Democritos, who believed in the existence of both atoms and empty space and to some extent Plato,
whose world of ideas also includes space as an ideal object, can be
regarded as representing an absolute view of space
while Aristotle with his opinion that space only epitomizes the place of the objects comes close to a relational
concept of space. During the age of Enlightenment the discussion about the status of space reached a new peak with
Descartes, Leibniz, Newton, Huygens, Berkeley and others. For Descartes \cite{descartes:1644} the central characteristic of bodies is
their extension which he identifies with space itself, i.e. they are not {\em in} space but constitute space by {\em
themselves}, there is no space devoid of bodies. Therefore, there are no external objects lending space its metric
properties, hence these properties must be given intrinsically. These characteristics are independent of the bodies and
thus represent an absolute feature. Leibniz \cite{erlichson:1967}, however, claimed that the relative position between
the bodies is sufficient for the definition of space and tried to support his view with the principle of sufficient
reason (principium rationis sufficientis): no location or orientation of objects in space is distinguished from any
other location or orientation, there is no reason for it. Hence all places are alike and it makes no sense to talk about
a certain location of bodies. In addition, Leibniz based his rejection of absolute space on another philosophical
principle, the identity of indiscernibles (principium identitatis indiscernibilium): when all objects in space are
displaced by the same amount, then this new situation cannot be discerned from the initial one and these two absolute
locations have no separable meaning. However, while Leibniz' relational point of view may be justified in kinematical
respects, it fails to defeat the dynamical arguments in favor of an absolute space brought up by Newton by means of his
famous bucket experiment. Based on the observational fact that a
vessel at rest filled with water and a similar vessel in
rotation differ in their shapes of the surface of the water, Newton was led to the notion of absolute motion: it is the
rotation with respect to absolute space which is responsible for the difference\footnote{Since the relative motion between the water
and the sides of the vessel is the same at the beginning of the experiment (vessel and water at rest) and at
the end (vessel and water in rotation), the relative velocity cannot account for the different shapes of the surface.}.
Therefore, though Newton clearly recognized that absolute space cannot be directly observed, he rejected the possibility
of basing physics on purely relational quantities.

For L. Euler the principles of mechanics which include the ideas of absolute time and absolute space are beyond any
doubt, since all consequences drawn from these principles are in perfect agreement with observation
\cite{euler:1748}. One is therefore led to consider these entities as real things rather than as imaginary and devoid
of any reality. Moreover, it would be absurd to maintain that pure imagination can serve as a foundation of the real
principles of mechanics. Also, depending on the absolute or relational view of space, the law of inertia would lead to
different predictions for the behavior of a body. To see this, Euler suggests to imagine
a body floating in water such that both the water
and the body are at rest. As long as there is no motion, the body remains in the same place in absolute space as well as
with regard to the water, which is in accordance with both views. However, when the water begins to flow, the body
should still remain at the same place in absolute space, while, according to the relational view, the body should
perfectly follow the movement of the water in order to preserve the neighborhood of the same particles of the water as
before. Although experience informs us that the body will be set in motion as soon as the water begins to flow, this
motion is caused by the particles hitting the body and it is thus an external force which puts the body into motion.
Without this force, the body will remain at rest both in the still and in the flowing water; therefore, in preserving
its state of motion, the body does not depend on the immediately surrounding matter and the idea of space as nothing but a
relation between bodies is thus untenable.

Further, Euler pointed out that the relational standpoint, i.e. regarding the heaven of fixed stars rather than absolute
space as the reference frame in which the laws of inertia are valid, would imply that the stars control the bodies
in their inertia. However, Euler considered this consequence as metaphysical and unacceptable.

In the transcendental philosophy of Kant \cite{kant:1781}, space and time are necessary prerequisites, i.e. a priori
conditions for establishing any kind of experience, since we cannot perceive objects without being able to represent
them spatially and temporally. Hence, space and time are absolute in the sense that they are pure forms of intuition and
cannot be modified by experience. This absolute space has to be distinguished from the empirical space which is the
epitome of all objects of experience and itself a matter of perception. Kant further notes that the mobility of objects,
i.e. the variation of the external proportions with respect to a given location cannot be recognized without recourse to
experience and that motion must thus be classified as an empirical term rather than as a notion of pure reason. Whether
a body is at rest or in uniform motion is not a {\em disjunctive} but an {\em alternative} decision\footnote{The first
refers to two objectively opposite predicates which are mutually exclusive, whereas the second applies to two in fact
objectively similar, yet subjectively opposite though not exclusive attributes.} and the state of motion is
therefore inherently indeterminate and a mere potential quality. In contrast, the circular (or any curvilinear) motion
of a body is a {\em true} attribute since it shows the existence of an external agitating force which prohibits the body
from moving along a straight line. The reverse, namely the motion of the empirical space and the body at rest is merely
kinematical and possesses no agitating force; hence in this case the decision whether the body or the space is in motion
is disjunctive. It is therefore not a matter of attitude whether the earth is rotating with the stars fixed or the other
way round; although this motion appears relative, the rotation of the earth is the true motion since it can be
demonstrated through the presence of inertial forces even in empty space \cite{kant:1786}.

In his effort to support the need of the idea of absolute space, Kant refers to the existence of incongruent objects
\cite{kant:1768}. These objects (e.g.\ a right and a left hand) agree in all metric properties and are thus equal with
respect to all internal relations, yet they can be distinguished from each other. Therefore, the reason for this
discrimination can only be due to external causes, i.e. to the proportion of the locations to space itself. According to
Kant, this inexplicability of the existence of incongruent objects on the basis of a relational view of space warrants
the introduction of absolute elements as agents of explanation.

Concerning the methodical analysis of the principles of dynamics, Lange \cite{lange:1885} considered the necessity of a
new and modern formulation of Newton's law of inertia as the ultimate ambition, since Newton's absolute space, which is
constituted by a series of unrecognizable absolute fixed points and which serves as a reference for the real locations
and motions of the physical bodies, can never establish the basis for an exact science. Therefore, it is essential to
find an appropriate substitute for the notion of absolute space. According to Lange, for absolute time this replacement
has already been managed: since there is no given motion in the world which can act as a suitable basis for chronometry,
the dynamical time is defined via the motion of a free body: two segments of time are considered equal when the
distances in space covered by a freely moving point are equal. In a similar way, because no material object in the
universe is adequate to serve in due form as a reference object for the law of inertia, Lange introduced a fundamental
coordinate system for dynamics which he called an inertial system and which is defined by means of three freely moving
points. Lange points to the fact that for up to three arbitrarily moving points one can always find a coordinate system
with respect to which these points move along straight lines. In general, however, this is no longer true for more than
three points. Hence, while the law of the constant direction of motion relies on a mere convention for three points
which are left to themselves, it is remarkable that it also holds for {\em any} number of points: the {\em physical}
condition of being unaffected thus results in the {\em kinematic} fact that there is a coordinate system in which an
{\em arbitrary number} of points are in uniform motion. The realization of such an inertial system can be accomplished
by three material bodies departing from a common origin and moving freely, i.e. without being exposed to a force; any
reference frame in which these masses move along straight lines then defines an inertial system. The law of inertia thus
becomes equivalent with the proposition that any other freely moving point mass likewise exhibits rectilinear motion
within such a system.

The most famous critique of Newtonian absolute space is due to Mach. In a critical and historical review\footnote{The
first edition of the {\em Mechanik} appeared in 1883.}
\cite{MA} he argued against the Newtonian interpretation of the bucket experiment by emphasizing that Newton's
conclusions were only true in a hypothetically empty universe and under the assumption that a physical system carries
its essential properties even if it is isolated in empty space. However, in the real universe filled with matter there
is an observable difference between the two states of the vessel: in
the case of the plane surface the vessel is at rest
with respect to the heaven of the fixed stars while in the case of a curved surface the bucket is rotating relative to
the stars. Therefore, this difference must be ascribed to the different states of motion with respect to the masses of
the universe rather than to absolute space. In other words, an abandonment of the rest of the world, as Newton did, is
not possible, and we always have to keep track of the whole world.

Mach was probably the first to point at dragging effects in the vicinity of rotating masses when he noted that the
bucket experiment only implies that the rotation of the water relative to the vessel does not induce any noticeable
centrifugal forces. However, such forces may be induced by rotation relative to the mass of the earth and the other
celestial bodies.

Mach insists that absolute motion and absolute space, i.e. motion and space in themselves, reside only in our minds and
cannot be revealed by experience, hence they are meaningless idle metaphysical concepts\footnote{Mach was an advocate of
the so-called empiriocritcism, a philosophical school which only regarded ``experience" as the source and last resort of
all cognition. This sensualism requires every proposition to be verifiable by reduction to sensations; all statements
which do not cope with this demand are considered as meaningless and thus condemned as metaphysical speculations.} and
must not be used in a scientific context. All our principles of mechanics are based on our experience about relative
locations and relative motions and we are not authorized to extend these principles beyond these limits.

Along the lines of Mach, B. and I. Friedlaender deny any difference between the mathematical and physical space, i.e.
between the space given by our intellectual power and that of perceptible phenomena and doubt the reality of absolute
motion~\cite{friedlaender:1896}. There is but relative motion and inertia should be explicable without recourse to
absolute elements. However, to avoid empty statements, it would either be necessary to find an improved form of the law
of inertia, or to demonstrate the inadequacy of the current conception of absolute motion experimentally. In particular
the phenomenon of centrifugal motion appears appropriate for an experimental solution of the problem: if the centrifugal
force occurring in a flywheel rotating against the earth is due to relative motion, then a similar force should arise
when the flywheel is at rest and the earth rotates with the same angular velocity in the opposite direction.
Accordingly, a centrifugal force effect, though much smaller, should be expected in fixed bodies close to rotating
massive flywheels. Thus a corresponding experiment using a torsion
balance was proposed in order to resolve the question
about the reality of absolute motion. However, the practical realization of this experiment in a rolling mill failed due
to uncontrollable disturbing influences.

Concerning the problem of the relativity of inertia, B. Friedlaender further notes that inertia, i.e. the resistance to
changes in the velocity, is not an internal property of a single body but rather a consequence of the influence of all
the other bodies of the universe. The law of inertia should thus be expressed in the following way: all masses tend to
maintain their state of motion of velocity and direction with respect to {\em each other}. However, the correct form of
the law of inertia has to await a unifying treatment of both inertia and gravitation, since both are effects of masses
on each other. The Friedlaenders have also pointed at a similarity between their concept of relative inertia and
induction effects in electromagnetism: just as a change in the magnitude of the current (or distance) will generate
induction effects, only changes in velocity will generate attractive (or repulsive) effects.

Machian ideas are also held by F\"oppl, the author of a well-known German introduction to Maxwell's theory of
electricity, who conjectures that the inertial systems are determined by the motions of all bodies of the universe. He
was inspired by Mach's observations on the physical significance of the law of inertia and on the concept of absolute
motion related with it. Proceeding from Mach's view, F\"oppl tried to add further considerations to the problem of
absolute and relative motion~\cite{foppl:1904}. Upon pointing out that the coincidence of the inertial frames --- which
are characterized by the absence of Coriolis forces --- with the heaven of the fixed stars cannot be regarded as
fortuitous, he ascribes this distinction to the influence of the masses of the stars. Thus one could ask about the law
which determines the orientation of the inertial system when the instantaneous positions and relative velocities of the
whole system are known. In case all bodies were at rest relative to each other, it is obvious from our experience that a
test mass would, when no forces act on it, describe a straight line with respect to the frame rigidly fixed with the
masses.

Now F\"oppl suggests to consider the case in which the bodies of the universe are divided into a large and a small group
with fixed distances within the bodies of each group. Any relative motion of the groups will alter the inertial system
and will induce a small motion of this system relative to the larger group due to the influence of the smaller group,
although it will still be almost at rest with respect to the large group. Therefore, one can fix the reference frame
with respect to the first group and take the influence of the smaller group into account by applying weak additional
forces of the relative motion, which the chosen system executes relative to the true inertial system. Once such a
decision is made, the Coriolis forces no longer appear as mere computational quantities arising from a coordinate
transformation, but rather as physically existing forces that are exerted by the masses of the smaller group and being
due to the motion of these masses with regard to the chosen reference frame.

One could start with the case that the second group consists only of a single mass and try to find the magnitude and
direction of the force at any point. This force will depend on the velocity of the single mass with respect to the
reference frame determined by the remaining bodies as well as on the distance to the single mass. When this problem has
been solved for a single body, the influence of a whole group of moving bodies can be obtained by the principle of
superposition.

For a realistic case, F\"oppl suggests to consider a reference frame determined by three suitably chosen stars. This
frame closely coincides with an inertial system since the constellation of the stars changes little within a time span
of several centuries. The small deviation from the true inertial frame can then be accounted for by applying Coriolis
forces which depend on the velocity of the masses of the universe relative to the chosen system.

F\"oppl holds the view that he has thus found a causal explanation for the existence of inertial frames: at any point
they are those frames in which all velocity-dependent forces arising from the masses of the universe balance each other.
With this, F\"oppl believes that he has obtained an adequate basis for the elaboration of the concept of absolute motion and
 constructed absolute space which appears in the law of inertia without abandoning the idea that all motion is
ultimately relative.

In F\"oppl's opinion the problems related with the law of inertia can only be solved upon assuming forces between the
bodies of the universe which depend on their velocities with respect to the inertial frames. Therefore, he proposes to
look for possible phenomena which allow to deduce the laws governing the velocity-dependent forces. However, when based
on astronomical observations, for F\"oppl this task does not seem promising because these forces, though distinct from
gravitational ones, might produce effects quite similar to them and it would be rather difficult to single out those
parts which are due to the velocity-dependent forces. Thus it appears more advisable to search for terrestrial phenomena involving
motions which might be influenced by the rotation of the earth. F\"oppl conducted a number of gyroscope experiments
by himself in order to detect phenomena which exhibit some deviating behavior of gyros from standard theory and
which might be induced by velocity-dependent forces; however, his attempts were without success.

A public demonstration of Foucault's pendulum experiment in Vienna caused Hofmann \cite{hofmann:1904} to
discuss critically the meaning of true and apparent motion and its connection with the notion of inertia. Hofmann
defines motion as any change in the location of a material body with respect to a sensually perceivable reference
frame; therefore, all motion is true and relative\footnote{Absolute motion of a body could be stated irrespective
of the existence of a second body. However, since motion can only be ascertained by virtue of sensual perception
and thus relies on the existence of observable objects of comparison, absolute motion is impossible due to the
absence of a reference object: it is only subjectively conceivable but not objectively perceivable.} and there is
no way to distinguish between true and apparent motion. Hence, contrary to common belief, Foucault's experiment
does not imply that the daily rotation of the earth is true while the observed motion of the sun about the earth
is only apparent --- the two views are not in conflict with each other, they are but different yet equal modes of
perception of an existing fact --- it implies something about the yet imperfectly realized law of inertia. Hofmann
considers the content of this law --- that a free body persists in its state of rectilinear and uniform motion ---
as incomplete since it gives no indication about the reference system in which the motion is rectilinear. But this
is essential, because a body can only be at rest or at constant velocity with respect to another body. According to
Hofmann, inertia must be defined such that every body with regard to all other existing bodies is subject to the law
of the conservation of the mutual state of motion and that its actual behavior is the resultant of all individual
influences. How this behavior depends on the mutual separations, masses and positions can only be explored by
experience and is not a matter of speculation. Further, if inertia is relative, i.e. a relation between the masses,
then the idea of a rectilinear inertial motion must be abandoned and replaced in general by curvilinear motion.
Another consequence of the relativity of inertia consists of the fact that for the determination of inertia the
simultaneous exertion of {\em all other masses} must be taken into account, in contrast to the determination of
the relative motion which has to be related to an arbitrary but {\em single} reference system.

Hofmann does not pretend to have developed a new theory of inertia, rather his intention is to pinpoint the deficiencies
of the actual law of inertia and to stimulate a reformulation of this law on a relational basis. In particular he
suggests a modification of the formulation of the kinetic energy ({\em vis viva}), based on the following argument:
consider two unequal masses $M$ and $m$ (with $M>m$) approaching each other with velocity $v$. Since the masses are
different, their kinetic energies are different as well and depend on whether $M$ or $m$ is chosen as the reference
frame. Now, upon an inelastic collision let a measuring device determine the kinetic energy of the masses, e.g. by
converting it into the energy of a compressed spring, this latter energy will be the same, irrespective of whether $M$
or $m$ has been considered at rest. Thus we arrive at the following: for two masses $M$ and $m$ in relative motion, the
{\em vis viva} of $M$ with respect to $m$ is the same as that of $m$ with respect to $M$. This conclusion, however,
appears to be incompatible with the standard formulation of the kinetic energy, which includes either $M$ or $m$ (but
not both) and should be rewritten in the form $E=kmMf(r)v^2$ in order to take {\em both} masses into account, where $k$
is a constant and $f(r)$ allows for a possible influence of the separation of the bodies.

Finally, Hofmann draws several interesting conclusions from the principle of the relativity of inertia.  The
centrifugal forces which appear for a body $K$ in rotation must be interpreted as an inertial relation between $K$ and
all external masses not participating in the rotation. A mass point of the rotating body $K$ tries to keep its position
relative to $K$ unchanged, while its inertia manifests itself in a continuation of its motion with respect to the
remaining bodies; both tendencies result in what we
call centrifugal forces. Consequently, the same forces must also appear when $K$ is conceived as fixed while all other
masses are rotating about it; therefore, in this latter case, the external masses should exhibit centrifugal phenomena
as well. However, while the individual mass points of these bodies obey the same two kinds of inertia as the mass points
of $K$, i.e. they exhibit a centrifugal force with respect to $K$ and hold their motion with respect to each other,
their resultant state of motion will turn out in favor of the overwhelming masses and hence will not show any measurable
centrifugal phenomena. This explains Newton's observations: although the tiny mass of the vessel will induce some
centrifugal forces in the water, the immutability of the position of the liquid elements is mainly due to the action of
inertia with regard to the much larger remaining masses. Hence this experiment does in no way demonstrate that relative
motion cannot produce centrifugal forces.

Hofmann seems to be the first who clearly stated that in a relational theory of inertia the kinetic energy should be
the sum over products of all pairs of masses rather than the sum of the contributions of individual masses. It is also
interesting to note that the work of Hofmann, though mentioned by Mach and Einstein, remained virtually unknown even
to the specialists who worked on the development of a new theory of inertia\footnote{Ten years after the publication
of Hofmann's booklet, in a paper by Reissner \cite{reissner:1914} many of the ideas anticipated by Hofmann appeared
and were believed to be new; although Reissner presumably knew of Hofmann's existence, he was obviously not aware of
his work.}.

\section{Mach's Principle and General Relativity} It is well known that Mach's ideas about the relativity of inertia
played an important role in the development of general relativity. Einstein aimed at an explanation of inertia which would
eliminate the privileged role of the class of inertial frames in classical mechanics, and which was based on the premise
that the results of measurements
should not depend on the choice of coordinates assigned to events \cite{einstein:1955}. Ironically, though general
relativity was intended to be based on relational concepts, contrary to its name it still contains absolute elements
and does not resolve the problem of the origin of inertia. This is already expressed in the calculations of the advance
of Mercury's argument of perihelion, which is referred to a coordinate system in which the gravitational potentials
assume certain boundary conditions at infinity. Empirically, this system coincides with the average system of the fixed
stars, however, this correspondence appears incidentally, since the presence of the distant masses did in no way enter
the calculations \cite{schrodinger:1925}.

Further, it has been shown that the relativity of all motion, i.e.\ the relativity of inertial accelerations, cannot be
maintained in general relativity. Starting from the assumption that inertial forces are of gravitational origin, neither
translational nor rotational acceleration is compatible with the relativistic theory of gravitation, since the latter is
based upon local Lorentz invariance and thus the notion of absolute motion cannot be avoided
\cite{mashhoon:1988}.
Moreover, the anti-Machian character of the field equations also shows up in solutions which allow for a curved
spacetime even in the absence of any matter \cite{taub:1951} as well as in solutions that exhibit an intrinsic rotation
of the matter with respect to the local inertial system \cite{godel:1949}.

These examples of the ontologically autonomous structure of space which completely determines the inertial properties
of test masses show the logical independence of Mach's principle from general relativity. Hence the idea that the metric
field is entirely governed by matter must be abandoned and Mach's principle can serve at best as a selection principle
for admissible cosmological solutions. In that case, however, an independent justification for the validity of such a
principle is required.

Mach's statement, given in the course of his analysis of Newton's bucket experiment, that the rotation
of the water with respect to the sides of the vessel does not induce noticeable centrifugal forces,
but that such forces may be induced by its rotation relative to the mass of the earth and other celestial bodies seems
to be the first clear hint of possible dragging effects near rotating bodies.

The calculation of dragging effects within the framework of general relativity was initiated by Thirring
\cite{TH} and Lense and Thirring \cite{LT}
who studied, based on the weak-field approximation, the dragging of inertial frames inside a slowly rotating mass shell
and outside a slowly rotating solid sphere, respectively. Later, upon investigating this effect for
arbitrarily large masses rotating slowly, it was shown that for compact masses, whose Schwarzschild radius approaches
the shell radius, the induced rotation approaches the rotation of the shell \cite{brill&cohen:1966}. This result could
explain the fact that local inertial frames do not rotate with respect to the distant matter. It is thus commonly
believed that dragging phenomena predicted by general relativity constitute the most direct manifestation of Machian
ideas in Einstein's theory of gravitation. It must be noted, however, that dragging phenomena also exhibit more or less
counterintuitive or anti-Machian features, depending on how one interprets Mach's writings
\cite{barbour&pfister:1995,ehlers&rindler:1971,R1,R1a}.

\section{Gravitomagnetism} In an attempt to unify electrostatics with
electrodynamics, i.e. Coulomb's law, Ampere's law and the laws of induction, Weber derived an expression for the force
between two charges in an arbitrary state of motion, depending on the relative velocity and acceleration as well as on
the propagation velocity of the electromagnetic action \cite{weber:1848,maxwell:1954}. However, this remarkable law of
force, which can be derived from a velocity-dependent potential and which refers only to relative motion\footnote{This
is distinctly different from the Lorentz force law which only holds with respect to an inertial frame.} and thereby
complies with the principle of action and reaction, is little known today mainly due to the work of Helmholtz who showed
that a force depending on distance and velocity does not conserve energy and thus erroneously claimed to have disproved
Weber. Although it became clear later that Helmholtz' proof does not apply to Weber's law, since the latter also
contains the relative acceleration of the particles, it nevertheless fell into oblivion in the twentieth century.
Anyway, Weber's electrodynamics was important concerning the first efforts to subsume the phenomena of gravitation under
electrodynamics. Upon extending Hamilton's method to velocity-dependent potentials, Holzm\"uller \cite{holzmuller:1870}
was among the first who studied the motion of a test particle that is attracted by a fixed center according to Weber's
electrodynamic law. He found that the trajectory is no longer closed but can be described by a slowly precessing
ellipse. In a similar way, upon assuming that the planetary motion is
governed by Weber's or Gauss' force law\footnote{These
two laws slightly differ from each other.} rather than by Newtonian
attraction alone, Tisserand \cite{tisserand:1872} showed
that the excess advance of Mercury's longitude of perihelion (after subtraction of all Newtonian perturbations of the
solar system bodies) could be explained by assuming that gravity propagates with a velocity of 6/7 of that of
light\footnote{This is based on Gauss' force law.}.

At the end of his book on electromagnetic theory, Heaviside \cite{heaviside:1894} also speculated about a
gravitational and electromagnetic analogy and worked out what could be considered in some sense as a low-velocity
and weak-field approximation to general relativity. Based on the similarity of
Newton's law with Coulomb's law and with Maxwell's theory in
mind, Heaviside constructed a set of field equations analogous to those of electrodynamics, thereby stating a finite
propagation speed of gravity and the existence of a gravitomagnetic field. He also noted that the analogy with
electrodynamics serves to emphasize the non-necessity of the assumption of an instantaneous action of matter upon matter
and that gravitational waves propagating through the ether may well move with the speed of light. Heaviside concluded
that the gravitational field of the sun, when moving with constant velocity $u$ through the ether will be modified by
terms of order $u^2/c^2$ and that there will be a slight weakening of the force in the line of motion and a slight
strengthening equatorially. Upon including a gravitomagnetic force, Heaviside noticed that the perturbations on the
orbit of the earth about the sun are too small as yet to be observed and are therefore not in disagreement with the
assumption that the speed of gravity may be the same as that of light.

Einstein's general relativity provided an elegant solution to the problem of the perihelion advance of Mercury that
involved a small gravitoelectric correction to the Newtonian attraction of the sun \cite{EI}. Soon after Einstein's
work, it was shown that the rotation of the sun generates a gravitomagnetic field within general relativity due to
mass currents that also leads to the precession of planetary orbits \cite{TH,LT,DS}. The effect turns out to be very
small and in the opposite sense of Mercury's excess precession. The origin of inertia and Mach's principle provided
the motivation for Thirring to investigate the gravitational field inside a rotating hollow shell \cite{TH,MHT}. If
the rotation of astronomical bodies is relative to the distant masses in the universe, then one might expect to
recover inertial forces inside a rotating hollow shell. Thirring showed the existence of a Coriolis-type force that
has been qualitatively interpreted as a Machian dragging effect. Moreover, Lense and Thirring~\cite{LT} gave a
general treatment of orbital precession due to the proper rotation of a central source. The Lense-Thirring precession
of planetary orbits due to the rotation of the sun is too small to be detectable at present; however, the effect may
be measurable for satellite orbits about the earth \cite{CI}.

The gravitomagnetic precession of a gyroscope due to a rotating source has been qualitatively interpreted in terms of
the dragging of the local inertial frames~\cite{SCH}. One of the main goals of the GP-B experiment, launched on April
20, 2004, and involving four superconducting spherical gyroscopes on board a drag-free satellite in polar orbit about
the earth, is the detection of the gravitomagnetic precession of the gyroscopes due to the rotation of the
earth~\cite{EV}.

Taking Mach's principle seriously either in terms of relativity of rotation~\cite{mashhoon:1988} or the existence of
inductive effects of dragging~\cite{R1,R1a,R2}, one can demonstrate quantitatively that the gravitomagnetic gyroscope
precession within general relativity is in conflict with Mach's principle. The conflict between general relativity
and Machian dragging can be clearly seen in the gravitomagnetic clock effect. Consider a circular equatorial geodesic
orbit about a central rotating mass and imagine two free test clocks on this orbit moving in opposite directions.
According to general relativity, the clock in {\em prograde} motion moves {\em slower} and takes {\em longer} than
the clock in retrograde motion to complete the orbit~\cite{M1,IO,M2,LI}.

Observations of the cosmic microwave background radiation have led to an extremely small upper limit on the frequency
of rotation of the universe. The presumed absence of rotation of the local inertial frame with respect to the frame of
the distant matter in the universe is often mentioned as evidence in favor of Mach's principle. However, this lack of
rotation is likely due to the circumstance that the universe is rather old and the rotational perturbations of standard
cosmological models decrease very rapidly with the expansion of the universe.

\section{Tact of the Natural Investigator}

In discussing Newton's views on time, space and motion in his great book on the science of mechanics~\cite{MA}, Mach
devoted a subsection (subsection 6, section VI, Chapter II of \cite{MA}) to the elaboration of Newton's fifth corollary
following the laws of motion in the {\em Principia}. In this corollary, Newton states the principle of relativity, which
is satisfied by his laws of motion based on absolute time and space. Commenting on Newton's conclusion that absolute
time and space are subject to the relativity principle, Mach remarks:
\begin{quote} ``\ldots In spite of his metaphysical liking for the
  absolute, Newton was correctly led by the {\em tact of the natural
    investigator}\ldots .''
\end{quote}

In view of the developments in theoretical physics in the past century, it is interesting to return to Mach's
outstanding critique of Newtonian mechanics and recognize the tact of the natural investigator in his own work. In the
penultimate subsection of the same section (subsection 11, section VI, Chapter II of \cite{MA}), Mach makes the
following observation:
\begin{quote}``\ldots We measure time by the angle of rotation of the
  earth, but could measure it just as well by the angle of rotation of
  any other planet. But, on that account, we would not believe that
  the {\em temporal} course of all physical phenomena would have to be
  disturbed if the earth or the distant planet referred to should
  suddenly experience an abrupt variation of angular velocity. We
  consider the dependence as not immediate, and consequently the
  temporal orientation as {\em external}. Nobody would believe that
  the chance disturbance --- say by an impact --- of one body in a system
  of uninfluenced bodies which are left to themselves and move
  uniformly in a straight line, where all the bodies combine to fix
  the system of co\"ordinates, will immediately cause a disturbance of
  the others as a consequence. The orientation is external here
  also. Although we must be very thankful for this, especially when it
  is purified from meaninglessness, still the natural investigator must
  feel the need of further insight --- of knowledge of the {\em immediate}
  connections, say, of the masses of the universe\ldots .''
\end{quote}

In this way, Mach arrives at the root of the epistemological difficulty in Newtonian mechanics: the internal
state of a Newtonian particle --- namely, its mass --- has no a priori connection with its external state in space
and time --- namely, its position and velocity. Space and time are basically different from their operational
definitions by means of masses \cite{LM}. That is, masses are simply ``placed'' in absolute space and time in
Newtonian mechanics, but have no organic connection with space and time.

As is well known, Mach's own resolution of this epistemological problem is to concentrate on the motion of masses
relative to each other. However, his basic analysis paves the way for other possible resolutions based on the
development of physics.

\section{Quantum Theory and Inertia}
Mach's analysis of the problem of motion was solely based on classical mechanics although the
principles of electrodynamics were already well established during the years of the later editions of his critique. In
an attempt to extend Mach's arguments to the motion of electromagnetic waves, Mashhoon \cite{mashhoon:1993} has pointed
out that the two kinds of motion, relative and absolute, are associated with particle and wave propagation,
respectively. On the one hand, we consider the velocity of a classical particle as relative, since a reference frame is
needed to specify the motion; in particular, there always exists a system in which the particle is at rest. On the other
hand,  due to Lorentz invariance, an electromagnetic wave cannot be at rest with respect to any inertial
observer; therefore, we do not need to specify a reference system for the propagation of light, i.e. its movement is
completely independent of the motion of inertial observers and it can thus be considered as absolute. Therefore,
classical mechanics and classical electrodynamics are concerned with two types of motion,
local particle and nonlocal wave motion, respectively. These are brought together in geometric optics, where the waves are replaced by
rays that can be treated in a similar way as classical point particles.

Concerning the motion of particles, as recognized by Mach \cite{MA}, there is no immediate connection between the
particle's {\em intrinsic} and {\em extrinsic} states, i.e. between its mass on the one hand and its position and
velocity on the other hand and the potentiality of relative motion can be viewed as the result of this independence. In
contrast, regarding the motion of electromagnetic waves, the state of the wave is directly related to its intrinsic
properties like period, wavelength and polarization, since they cannot be separated from the extrinsic wave function and
the impossibility of an observer comoving with the wave, i.e. its absolute motion, may be attributed to this coupling of
the intrinsic and extrinsic states \cite{mashhoon:1988}. This further suggests the hypothesis that an electromagnetic
radiation field cannot stand still with respect to {\em any} observer \cite{mashhoon:1988,mashhoon:1993}.

Upon extending these considerations to quantum physics, it is expected that the propagation of matter waves is also
independent of the motion of the observer. Indeed, in order to stay at
rest with respect to a given system, a particle is required to occupy a definite position in space and simultaneously exhibit zero momentum. Due to its wave nature
these requirements cannot be met by a quantum particle, thus its motion is absolute and it is impossible for an inertial
observer to be comoving with a matter wave; this is feasible only in the classical limit ($m\rightarrow\infty$). Hence
the uncertainty principle via the basic quantum condition in the
Heisenberg picture, $[x_j,v_k]=i\hbar\delta_{jk}/m$, constitutes through Planck's constant a direct connection
between the position $\bf x$ and the velocity $\bf v$ of a particle and its inertial mass $m$ and this
relation, just as in the electrodynamic case, eventually entails absolute motion. With increasing mass, the wave
aspect becomes negligible and in the limit $m\rightarrow\infty$ the classical behavior of the particle is recovered.
Thus, due to the wave nature of a quantum particle, its motion might
be considered in terms of the complementary notions of absolute
and relative movements in correspondence with the duality of waves and particles \cite{mashhoon:1988,mashhoon:1993}.

Based on the assumption that the above considerations also hold for accelerated observers, it may be concluded that
electromagnetic and quantum phenomena can be used to determine the absolute state of motion since they establish an
absolute frame of reference subject to Lorentz invariance. In particular, the detection of the earth's rotation using
superfluid helium~\cite{schwab et al:1997} could be interpreted in terms of the fact that once the fluid acquires its
coherence the wave nature becomes dominant and thus exhibits absolute motion~\cite{mashhoon:1993}. Finally it should
be noted that the inertia of a quantum particle is determined by its inertial mass as well as its spin, since these
characterize the particle's wave function.

\section{Discussion}

Mach's profound critique of the foundations of Newtonian mechanics played a key role in Einstein's development
of general relativity. Mach's principle has also guided other developments in gravitation theory such as the
scalar-tensor theories~\cite{JO,BD}. It has inspired interesting experiments, such as the Hughes-Drever experiment
on the local isotropy of space~\cite{CS,HU,DR} and continues to be of current interest~\cite{HN,FU}.

Mach identified the essential epistemological shortcoming of the Newtonian foundations of physics, namely, that the
intrinsic state of a particle in Newtonian mechanics, i.e. its mass, has no immediate connection with its extrinsic
state in space and time, i.e. its position and velocity. Mach's observation can be re-stated in terms of the a priori
independence of position (${\bf x}$) and momentum (${\bf p}$) of a particle in Newtonian mechanics. Let us note that
this deficiency has been overcome in the quantum theory by the fundamental quantum condition
$[x_j,p_k]=i\hbar \delta_{jk}$.

In quantum mechanics, mass and spin are both measures of inertia. Therefore, there are inertial effects proportional
to Planck's constant, such as the spin-rotation coupling, that are due to the inertia of intrinsic spin
\cite{mashhoon:1986}.

\newpage
\section*{Acknowledgments}

BM is grateful to Friedrich Hehl for helpful discussions and
correspondence. HL is indebted to B. Besser for his help in
obtaining some of the original papers.


\end{document}